\documentclass{article}

\usepackage[backend=bibtex]{biblatex}

\usepackage[pdftex]{graphicx}
\usepackage{amsmath}
\usepackage{amsfonts}
\usepackage{algorithmicx}
\usepackage{algpseudocode}
\usepackage{algorithm}
\usepackage{array}
\usepackage{color}
\usepackage{hyperref}
\usepackage{url}
\usepackage{fullpage}
\usepackage[T1]{fontenc}

\bibliography{ref}

\newcommand{\paran}[1]{\left( #1 \right)}
\newcommand{\nonadap}[3]{\texttt{NON-ADAPTIVE}\paran{#1, #2, #3}}

\title{Beating the Multiplicative Weights Update Algorithm} 
\author{Abhinav Aggarwal, Jos\'e Abel Castellanos Joo, Diksha Gupta \\ University of New Mexico, Albuquerque, NM}

\date{April 25, 2017}

\begin{document}
\maketitle

\begin{abstract}
Multiplicative weights update algorithms have been used extensively in designing iterative algorithms for many computational tasks. The core idea is to maintain a distribution over a set of \textit{experts} and update this distribution in an online fashion based on the parameters of the underlying optimization problem. In this report, we study the behavior of a special MWU algorithm used for generating a global coin flip in the presence of an adversary that tampers the experts' advice. Specifically, we focus our attention on two adversarial strategies: (1) non-adaptive, in which the adversary chooses a fixed set of experts a priori and corrupts their advice in each round; and (2) adaptive, in which this set is chosen as the rounds of the algorithm progress. We formulate these adversarial strategies as being greedy in terms of trying to maximize the share of the corrupted experts in the final weighted advice the MWU computes and provide the underlying optimization problem that needs to be solved to achieve this goal. We provide empirical results to show that in the presence of either of the above adversaries, the MWU algorithm takes $\mathcal{O}(n)$ rounds in expectation to produce the desired output. This result compares well with the current state of the art of $\mathcal{O}(n^3)$ for the general Byzantine consensus problem. Finally, we briefly discuss the extension of these adversarial strategies for a general MWU algorithm and provide an outline for the framework in that setting.
\end{abstract}

\newcommand{\abhinav}[1]{(\textbf{\texttt{ABHINAV : }#1})}

\section{Introduction}

Multiplicative weights update (MWU) algorithms form a general class of iterative algorithms that have a large number of applications in the field of computer science, specifically in areas related to but not limited to linear and semi-definite programming. Based on the excellent survey by Arora et. al~\cite{arora2012multiplicative}, such algorithms generally maintain a distribution over a certain set, which is then updated iteratively by scaling with respect to a given payoff function associated with the elements of the set. The idea is to set the parameters of this iterative scaling in a way that after a certain number of rounds, the algorithm converges to a desired function of the elements in the set. Some popular applications of this technique include the Adaboost algorithm for obtaining non-linear classifiers~\cite{freund1995desicion}, game playing in economic environments~\cite{freund1999adaptive}, portfolio management~\cite{helmbold1998line} etc. \\

In this report, we play the role of a devil's advocate to study how an MWU algorithm performs in the presence of a greedy adversary who tries to delay the time it takes for the algorithm to converge to its required goal. In particular, we assume that $n$ experts are giving advice in each round, with respect to which the algorithm is trying to make predictions as close to the best expert as it can. The adversary then tries to set the advice of some of the experts in a way that maximizes the total weight of these experts at the end of each round and makes the algorithm predict in a direction opposite to a chosen hidden direction, for as long as it is possible for him to do so. We study how the adversary decides what experts to capture and how to set their advice by formulating an optimization problem that he needs to solve in every round. We assume that the adversary has exponentially large computational power and has read-only access to the advice of the good experts. We then study the behavior of the MWU algorithm against this adversary and verify that indeed it takes longer for the algorithm to converge compared to when there was no adversary. We provide an empirical evidence in the favor of our claim that the global coin flip MWU algorithm terminates within $\mathcal{O}(n)$ rounds with the desired output.\\

 The motivation behind trying to maximize the weight of the corrupted experts is that by doing so, the adversary is able to use his experts to have a large share in the weighted advice that the MWU uses to compute its output. However, we do not make any claims about our adversarial strategy being the strongest possible. Many other adversarial strategies can delay the MWU algorithm as well. Nevertheless, we compare our results with the best known bound of expected $\mathcal{O}(n^3)$ time for byzantine consensus as provided by King and Saia in~\cite{king2016byzantine}.
 
\subsection{MWU setup}
The MWU that we consider for our experiments and theoretical analysis is a simple iterative algorithm to generate a global coin flip. We have $n$ experts, out of which $\tau < n$ are controlled by the adversary. If the adversary is adaptive, he chooses these experts as the rounds proceed, else he chooses this set of experts a priori.\footnote{As we will explain later, in each round, the adaptive adversary has an additional advantage of choosing up to $\sqrt{n}$ of good experts to flip advice of, apart from these corrupted experts.} Each of the remaining $n-\tau$ experts output uniformly at random from $\{-1,+1\}$ in each round, while the corrupted experts have their advices set by the adversary. In the beginning of every round, the adversary chooses a hidden direction in $\{-1,+1\}$ and tries to set the advice of his experts in a way that the weighted advice of all the experts, when passed to the function $sgn(x) = +1 \text{ if}\ x\geq 0 \text{ and } -1 \text{ otherwise}$, is in the direction opposite to the chosen hidden direction. The goal of the MWU algorithm is to weight the experts in a way that eventually, this weighted advice is in the hidden direction that the adversary chooses, irrespective of how the corrupted experts advice. The hidden direction is revealed once the algorithm produces its output for the ongoing round. The gain associated with each expert, which is used to update their weights, is $1$ if the expert advices in the same direction as the hidden direction and $-1$ otherwise. We assume that the weight update factor $\eta < 1/2$ is appropriately set to minimize the time to convergence.
\subsection{Notation and Terminology}
In compliance with standard notation and terminology when using MWU algorithms, we call the experts taken over by the adversary as being \emph{corrupted}, and refer to all others as being \emph{good}. In a given round $r$, we denote by $a_i^{(r)}$ the advice of expert $i$ in this round, $w_i^{(r)}$ the weight of expert $i$, $z^{(r)}$ the output of the MWU algorithm and $m_i^{(r)}$ as the gain of expert $i$ in this round. Additionally, we denote by $N$ the set of all experts and $K^{(r)}$ the set of corrupted experts in round $r$. We denote the hidden direction chosen by the adversary in round $r$ by $d^{(r)}$. As mentioned before, we denote by $sgn(x)$ a function that returns $+1$ if $x \geq 0$ and $-1$ otherwise.
\subsection{Experimental Setup}
We verify that the adversarial strategies as presented delay the runtime of the MWU algorithm through experiments. We coded these strategies in Python environment and ran our experiments on a Mackbook Pro/Air. The plots were obtained using the modules provided by the numpy library~\cite{oliphant2006guide}. 
\subsection{Paper organization} 
This report is divided into four main sections. Section~\ref{sec:nonadap} discusses the non-adaptive adversarial strategy by first formulating the underlying optimization problem and then providing the plots we obtained upon running experiments using this strategy. Section~\ref{sec:adap} discusses the adaptive adversarial strategy in a similar manner. Section~\ref{sec:generalization} briefly discusses how the concepts in this paper can be extended to a general MWU algorithm. We refer to this problem as the \emph{dual} of the MWU algorithm. Finally, we conclude the results of this report in the last section.

\section{Non-Adaptive Adversary}
\label{sec:nonadap}
We first consider the case when the adversary is non-adaptive, i.e. he has chosen a fixed set of $\tau$ experts a priori and he will set the advice of these experts in each round. No other experts will be taken over by the adversary during the course of the algorithm. We assume $\tau = \Theta(n)$. However, since the number of corrupted experts is linear in the total number of experts and the set $K$ of these corrupted experts is unknown in advance, the adversary can bias the output of the algorithm in any direction he wants if he has a large share in the weighted advice that the MWU computes. Hence, in any given round $r$, if the hidden direction chosen by the adversary is $d^{(r)} \in \{-1,+1\}$, then he tries to compute a subset $J \subseteq K$ of experts whose advice he will set to $d^{(r)}$. The experts in $K \setminus J$ will then advice $-d^{(r)}$ in this round and all other good experts will advice uniformly at random. Recall that the goal of the adversary is to make the weighted advice point in the direction $-d^{(r)}$, while ensuring that the total weight of the corrupted experts is maximized at the end of round $r$. This can be formulated in the form of the following optimization problem that the adversary solves in round $r$.
\begin{equation}
\label{eq:optEq1}
	\max_{J \subseteq K} \left( \sum_{i \in J} (1+\eta)w_i^{(r)} + \sum_{i \in K \setminus J} (1-\eta)w_i^{(r)} \right)
\end{equation}
subject to
\begin{equation}
\label{eq:constraints1}
	sgn \left( \sum_{i \in J}d^{(r)} w_i^{(r)} - \sum_{i \in K \setminus J}d^{(r)} w_i^{(r)} + \sum_{i \in N \setminus K} a_i^{(r)}w_i^{(r)} \right) + d^{(r)} = 0
\end{equation}
We can further simplify this problem by some algebraic manipulations. Observe that in a fixed round $r$, the sum of the weights of the corrupted experts, $\sum_{i \in K}w_i^{(r)}$, the weighted advice of the good experts, $\sum_{i \in N \setminus K}a_i^{(r)} w_i^{(r)}$ and the weight update factor $\eta$ are also fixed, since the adversary chooses $J$ only after seeing the advice of the good experts. \\

Let $f_r(J) = \sum_{i \in J} w_i^{(r)} - \sum_{i \in K \setminus J}w_i^{(r)}$. Then the above optimization problem is equivalent to maximizing $f_r(J)$ subject to the constraint that 
\begin{equation}
\label{eq:constraints2}
	sgn \left( d^{(r)} f_r(J) + \sum_{i \in N \setminus K} a_i^{(r)}w_i^{(r)} \right) + d^{(r)} = 0.
\end{equation} 

\subsection{Formulating the problem as $0-1$ integer linear program}
We first show that the problem in Eq.~\ref{eq:constraints2} can be formulated as a special case of the 0-1 integer linear program. For $i \in K$, let $X_i = 1$ if $i \in J$ and $0$ otherwise. Then $f_r(J)$ can be written as a function of $X = \{ X_i \mid i \in K \}$ as $\sum_{i \in K}(2X_i - 1)w_i^{(r)}$. Since $\sum_{i \in K}w_i^{(r)}$ is fixed in round $r$, maximizing $f_r(J)$ is the same as maximizing $g_r(X) = \sum_{i \in K}X_i w_i^{(r)}$, which is a linear function in $X_i$'s. This maximization will be subject to the constraints that 
\begin{equation}
\label{eq:constraints3}
	sgn \left( d^{(r)} \sum_{i \in K}(2X_i - 1)w_i^{(r)} + \sum_{i \in N \setminus K} a_i^{(r)}w_i^{(r)} \right) + d^{(r)} = 0.
\end{equation}
and that each $X_i \in \{0,1\}$. Note that since $d^{(r)}$ is known to the adversary, this constraint is actually a linear constraint on $X_i$'s. Hence, we have formulated the optimization problem that the adversary solves as an instance of the 0-1 integer linear program. Since this problem is NP-Complete in general, we may need to approximate the solution. However, we show that this particular instance can be solved using dynamic programming in the following subsection.

\subsection{Dynamic programming solution}
Without loss of generality, assume $d^{(r)} = 1$. The case when $d^{(r)} = -1$ can be solved similarly. Then Eq.~\ref{eq:constraints3} simplifies to the following inequality.
\begin{equation}
\label{eq:constraints4}
	\sum_{i \in K}X_iw_i^{(r)} < c_r
\end{equation} 
where $c_r = \frac{1}{2}\left( \sum_{i \in K}w_i^{(r)} - \sum_{i \in N \setminus K} a_i^{(r)}w_i^{(r)} \right)$ is fixed for round $r$. Note that this is just the constraint that $g_r(X) < c_r$. Thus, we need to maximize $g_r(X)$ subject to the constraint that it never grows to more than $c_r$ and that each $X_i \in \{0,1\}$. \\

For $1 \leq m \leq \tau$ and $c \geq 0$, define $M(m,c) = \max_{X \in \{0,1 \}^{|K|}} \sum_{i=1}^m X_i w_i^{(r)}$ subject to $\sum_{i=1}^m X_i w_i^{(r)} < c_r$. Then, it must be the case that $M(1,c) = w_1^{(r)}$ if $w_1^{(r)} < c$ and $0$ otherwise. Since all expert weights are positive, we let $M(m,c) = -\infty$ if $c < 0$. Finally, for $m > 1$, we have $M(m,c) = \max \left \{ M(m-1,c - w_m^{(r)})+w_m^{(r)}, M(m-1,c) \right \}$. Note that this dynamic program runs in time $\mathcal{O}(\tau c_r) = \mathcal{O}(\tau)$ in round $r$. \\

Thus, in each round, the adversary, after fixing $d^{(r)}$, solves the corresponding optimization problem to determine how to set the advice of the corrupted experts and in the process, ensures he keeps a high weight to himself at the end of each round. We provide empirical results of the performance of our MWU against this adversary in the next section. However, before closing this subsection, we must point out that if the adversary was adaptive, he solves a very different optimization problem. In particular, he needs to determine first if the corrupted experts can affect the outcome or not. If yes, he need not capture any new people, but in case they aren't able to, then he will capture people according to some strategy. We explore the adaptive adversary in the next section.

\subsection{Empirical Results}
In this section, we present some empirical results that show how the MWU performs against the adversarial strategy described above. Each plot was obtained by fixing $\tau = n/10$, and for each point plotted, an average over 20 runs was taken. The value of $\eta$ was set to be $\sqrt{\log n/ n}$ for the MWU algorithm. \\

\begin{figure}
\centering
\includegraphics[width=0.70\textwidth]{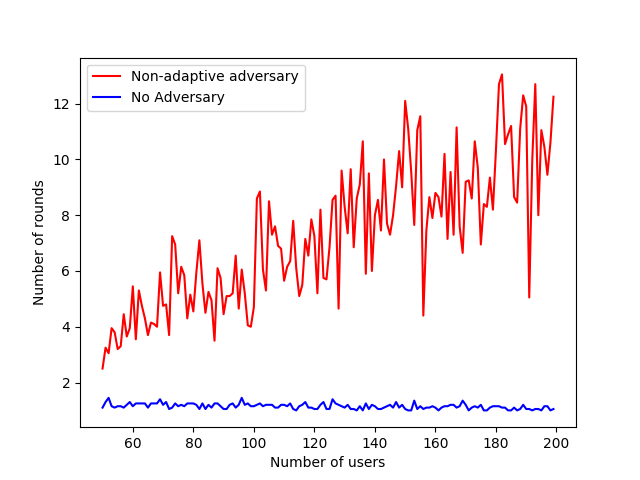}
\caption{Plot of the number of rounds before the MWU outputs in the hidden direction vs. the total number of experts. For each value of the number of experts, the corresponding number of rounds plotted is the average of running the experiment 20 times. Furthermore, the plot was obtained by keeping the fraction of the corrupted experts to be a tenth of the total number of experts.}	
\label{img:MWUtimevsnodes}
\end{figure}

Our first plot is in Fig.~\ref{img:MWUtimevsnodes}. It shows that in the absence of any corrupted nodes, the MWU algorithm is able to output in the hidden direction in fewer than 2 rounds in expectation. This is consistent with what should really happen, since if all experts advice uniformly at random, then the hidden direction, which is chosen before the experts present their advice, will be output in a single round, in expectation.\\

When the adversary has captured $\tau$ experts, the number of rounds now increases as $n$, and hence, the number of corrupted experts increase.  The plot hints towards a linear increase in the running time, which suggests that in the presence of a greedy non-adaptive adversary (as we described previously), the expected number of rounds before the MWU outputs the hidden direction is $O(n)$. Of course, this is only empirical evidence in the favor of this claim. However, the important point to note here is that the greedy strategy deployed by the adversary seems to have succeeded in making the algorithm run for a larger number of rounds, since the adversary controls a large fraction of the total weight for as long as he can.  \\

\begin{figure}
\centering
\includegraphics[width=0.70\textwidth]{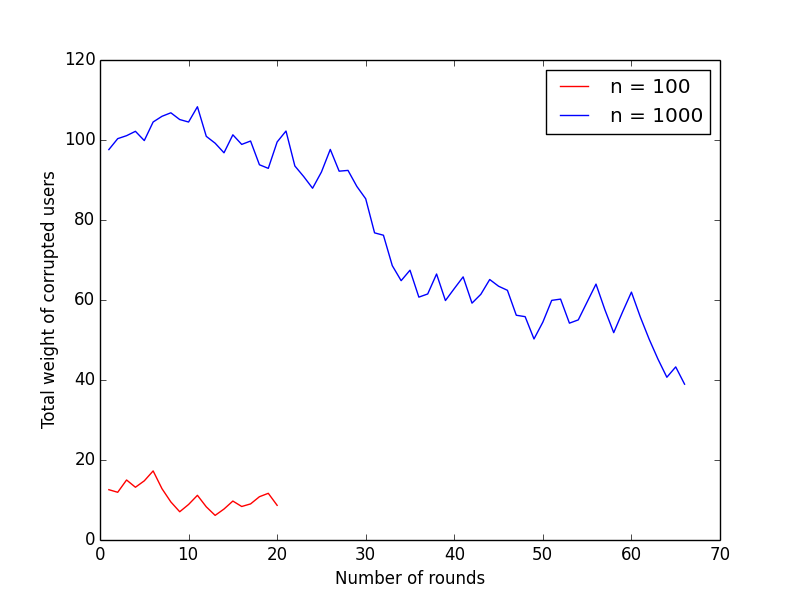}
\caption{Plot showing how the total weight of the corrupted experts varies over the rounds of the MWU. For each round, the corresponding total weight of corrupted experts plotted is the average of running the experiment 20 times. Furthermore, the plot was obtained by keeping the fraction of the corrupted experts to be a tenth of the total number of experts.}	
\label{img:MWUweightofcorruptedusers}
\end{figure}

Yet another plot which we think is worth studying is in Fig.~\ref{img:MWUweightofcorruptedusers}. It shows that although the adversary controls a large fraction of the total weight of the experts, this advantage with him declines as the rounds advance (after a certain threshold). For $n = 1000$ experts, the plot suggests that after about 11 rounds, the total weight of the corrupted experts starts decreasing. Although this decrease is not monotonic (partially because of the way MWU works and the fact that the good experts advice uniformly at random), there is an overall decrease in this quantity as the rounds progress. This certainly matches our intuition of what must happen for the MWU to eventually terminate. If the total weight of the corrupted experts always increased with the rounds, then the adversary would control an increasing fraction of the total weight, and hence, have a higher \emph{impact} on the outcome of the MWU. In this case, the MWU would have never been able to output in the hidden direction because the adversary would drive the output away from the hidden direction with more force every round. Hence, an eventual decrease is necessary for termination. \\

However, the total weight with the adversary is also expected to first increase for few rounds before starting to eventually decrease (which is also the case in the plots). We claim that this is because of the greedy strategy. The adversary tries to set the advice of his experts in a way that he maximizes their total weight at the end of each round. He is able to do so fruitfully for some rounds. But at the same time, each good expert predicts in the correct direction in expected half the rounds so far, and hence, his weight is also changing. For each good expert that predicts in the correct direction for more than half the total rounds (even 1 more than half will suffice), the net weight for him is more than $1$ (which was his initial weight when the MWU began running). We know there is a constant fraction of such experts in expectation. Thus, when enough rounds have passed so that these good experts are weighted high enough, there is only so much the adversary can do with his experts to counter the effect. The more experts the adversary forces to advice in the opposite direction, the lesser is the total weight he will have at the end of the current round. Hence, there will come a point after which the good experts start taking over and their total weight starts increasing. Eventually, they will win the race and the MWU will predict in the hidden direction. 

\section{Adaptive Adversary}
\label{sec:adap}
We now consider the case when the adversary is adaptive. More specifically, the adversary can choose up to $\tau$ experts to corrupt, but he need not make this choice a priori. As the rounds of the MWU progress, the adversary can choose to take over an expert and then corrupt his advice in the subsequent rounds. Additionally, in each round, the adversary can choose up to $\sqrt{n}$ good experts and flip their advice. This set can be different in every round. We refer to these experts as being \emph{volatile}. Although this adversary is not strictly a byzantine adversary in full generality, we will sometimes refer to him as a byzantine adversary for our discussion. To formulate the optimization problem for such an adversary, we use the discussion of the non-adaptive adversary from the previous section. However, unlike the results there, we prove that the problem for the adaptive adversary requires exponential time to solve and hence, we require the adversary to be able to perform exponential amount of computation locally. \\

From the previous section, recall the non-adaptive adversary, which when corrupts a fixed set $K \in N$ of experts in some round $r$ with respect to the chosen hidden direction $d^{(r)}$, solves the following optimization problem to determine which set of experts should advice in the direction $d^{(r)}$ and which ones should advice $-d^{(r)}$. For $i \in K$, let $X_i = 1$ if the expert $i$ advices in the direction $d^{(r)}$, and $0$ otherwise.

\begin{equation}
\label{eq:nonadap_opt}
	\max \sum_{i \in K} X_i w_i^{(r)} \quad \text{subject to} \quad \sum_{i \in K} X_i w_i^{(r)} < c_r
\end{equation}
where $c_r$ is a constant that depends on $r$ and $d^{(r)}$. Now, let $\nonadap{K}{r}{d^{(r)}}$ denote the function that solves this optimization problem above. More precisely, $\nonadap{K}{r}{d^{(r)}}$ first computes a vector $X \in \{0,1\}^{|K|}$ that maximizes $\sum_{i \in K} X_i w_i^{(r)}$ subject to $\sum_{i \in K} X_i w_i^{(r)} < c_r$, and then returns the total weight of corrupted experts at the end of round $r$ when the adversary tries to make the MWU algorithm output in $-d^{(r)}$ direction. \\

To make the math easier, we return the total weight only when some corrupted expert advices in $d^{(r)}$ direction. The idea here is that if the adversary realizes that all experts need to advice opposite to the hidden direction, then his total weight must be reduced by the maximum amount at the end of this round. More importantly, it is possible that the algorithm has weighted the good experts by this round in a way that now no matter what the adversary does, the output will be in the hidden direction. Thus, we assume pessimistically that when this is the case, the adversary sees this as a bad round for himself and hence, make $\nonadap{K}{r}{d^{(r)}}$ return $0$ in this case. For a proposition $p$, let $\mathbb{I}(p) = 1$ if $p$ is true and $0$ otherwise. Then, we get the following.
\begin{equation}
\label{eq:nonadap}
	\nonadap{K}{r}{d^{(r)}} = \sum_{i \in K} w_i^{(r+1)}\mathbb{I}\paran{\sum_{i \in K}X_i > 0}
\end{equation} where $X_i \in \{0,1\}$ is chosen such that $\sum_{i \in K}X_i w_i^{(r)}$ is maximized, while keeping it below $c_r$.

\subsection{The optimization problem}

We will now use this strategy for a non-adaptive adversary to design a strategy for a greedy-adaptive adversary. Again, by greedy, we mean that the adversary tries to maximize the weight of the corrupted experts at the end of every round. We assume that the adversary tries to capture as small a set of experts as possible in any given round, which is just enough to make the MWU algorithm output opposite to the hidden direction. We formalize this strategy in the form of an optimization problem as follows. \\

Fix a round $r$ of the MWU algorithm. Let $K^{(r)}$ be the set of corrupted experts at the beginning of this round. We assume that $K^{(1)} = \emptyset$, i.e. all experts are good to begin with. For a given $m \geq 1$, we denote by $N \setminus K^{(r)} \choose{m}$ the set of all subsets of $m$ experts from $N \setminus K^{(r)}$. Recall that once $r$ is fixed, so are $K^{(r)}, d^{(r)}, w_i^{(r)}, a_i^{(r)}$ and $c_r$ at the time when the adversary chooses what to do in this round. \\

\subsection{Corrupted experts} Let us first focus only on the corrupted experts. In the round $r$, if the adversary has already taken over $|K^{(r)}|$ experts, he can corrupt a maximum of $\tau - |K^{(r)}|$ more. However, since he tries to minimize the number of experts that he takes over, he first checks to see if he can set the advice of the currently corrupted $K^{(r)}$ experts to make the MWU algorithm output opposite to the hidden direction. If this is possible, then he need not take over any more experts. This is equivalent to looking at the output of $\nonadap{K^{(r)}}{r}{d^{(r)}}$. If the output of this is $0$, then since the weights are always positive, it must be the case that all experts in $K^{(r)}$ had to advice in $-d^{(r)}$ direction. When this happens, either the adversary still wins or it is the case that experts in $K^{(r)}$ are insufficient to establish the adversary's goal. If the latter is true, the adversary has no option but to capture more people (if allowed by $\tau$), but in the former case, we use the fact that it possible to capture some more people and prevent everyone from predicting in the direction opposite to the hidden one, because when this happens, everyone's weight reduces. Thus, when $\nonadap{K^{(r)}}{r}{d^{(r)}}$ returns $0$, our adversary hunts for new experts to capture. \\

However, when $\nonadap{K^{(r)}}{r}{d^{(r)}}$ is non zero, then the adversary is assured that the current set of experts can be used to set advices in a way that the MWU outputs against the hidden direction. Hence, he does not capture any new expert in this case. He acts similar to a non-adaptive adversary in this round to determine an optimal set of experts that will advice in $d^{(r)}$ so that the total weight of corrupted experts is maximized at the end. Thus, from now on, we assume that $r$ is the index of a round in which the function $\nonadap{K^{(r)}}{r}{d^{(r)}}$ returns $0$, indicating a need for the adversary to capture more experts. Furthermore, we assume $\tau - |K^{(r)}| > 0$, or else the adversary is unable to capture more experts and the algorithm terminates soon after round $r$. \\

We start by looking at what happens if the adversary captures $m$ experts from the set of good experts $N \setminus K^{(r)}$, for $1 \leq m \leq \tau - |K^{(r)}|$. Given such an $m$, the problem is to find a set $J_m \subseteq N \setminus K^{(r)}$ such that $|J_m| = m$ and $\nonadap{K^{(r)} \cup J_m}{r}{d^{(r)}} \neq 0$. Furthermore, since we want a greedy adversary who tries to maximize the weight with himself, we must find a $J_m$ that maximizes the return value of $\nonadap{K^{(r)} \cup J_m}{r}{d^{(r)}}$. Hence, when $m$ is fixed, the best adversarial strategy is to take over experts in the set $J^*_m$ such that the following holds.
\begin{equation*}
	J^*_m = \arg\!\max_{J \in {N \setminus K^{(r)} \choose{m}}} \nonadap{K^{(r)} \cup J}{r}{d^{(r)}}
\end{equation*}
Clearly, we want to choose the smallest $m$ for which $\nonadap{K^{(r)} \cup J^*_m}{r}{d^{(r)}}$ is non-zero, because the number of experts that need to be captured is to be kept as low as possible. If no such $m$ exists, then the MWU algorithm must terminate $\Theta(1)$ rounds after round $r$. Note that the search space for this problem is exponential in the number of good experts in round $r$. Since $\tau = \Theta(n)$, the search space is $\mathcal{O}(n^{\tau - |K^{(r)}|})$. However, simple algebra shows that $\nonadap{K^{(r)} \cup J}{r}{d^{(r)}}$ is not convex for any $m$ and hence, the problem is non-trivial to solve exactly.   

\subsection{Volatile experts}
Once the adversary realizes that he needs to capture more experts, he has two choices. Either he chooses experts according to the previous subsection, or he can act smart and use his additional power of flipping the advice of up to $\sqrt{n}$ good experts. Only when the latter fails to help, will he then capture new experts. Note that this additional power prevents a trivial strategy to produce the global coin, which just uses the advice of the $i^{th}$ expert in round $i$. \\

To select a set $V^{(r)} \in N \setminus K^{(r)}$ of $|V^{(r)}| = v \leq \sqrt{n}$ good experts to flip advice of in round $r$, the adversary must check if $\nonadap{K^{(r)} \cup V^{(r)}}{r}{d^{(r)}}$ returns a non-zero value. Similar to the case of corrupted experts, this set $V^{(r)}$ must be chosen such that $v$ is as small as possible. Thus, combining this selection of volatile experts with that of the corrupted experts, in round $r$, the adversary performs the following steps. \\

Let $0 \leq m \leq \tau - |K^{(r)}|$.
\begin{enumerate}
	\item Find $J_m \in {N \setminus K^{(r)} \choose{m}}$ such that $\nonadap{K^{(r)} \cup J_m}{r}{d^{(r)}} \neq 0$. If multiple such sets exist, choose the one with the highest total expert weight and return. 
	\item If no such $J_m$ exists, then for each $J \in {N \setminus K^{(r)} \choose{m}}$, find $V_J^{(r)} \in {N \setminus \paran{K^{(r)} \cup J} \choose{\sqrt{n}}}$ such that $\nonadap{K^{(r)} \cup J \cup V_J^{(r)}}{r}{d^{(r)}} \neq 0$. If multiple choices exist, choose one arbitrarily. Else, if no such $V_J^{(r)}$ exists for any $J$, set $m \gets m + 1$ and repeat from the first step.
	\item If $m > \tau - |K^{(r)}|$, then do nothing and return.
\end{enumerate}
These three steps can be summarized as follows. The adversary tries to compute a set $S^*_{m}$ as a solution to the following optimization problem.
\begin{equation}
	\max_{J \in {N \setminus K^{(r)} \choose{m}}}\  \min_{0 \leq v \leq \sqrt{n}}\ \max_{V \in {N \setminus \paran{K^{(r)} \cup J} \choose{v}}} \nonadap{K^{(r)} \cup J \cup V}{r}{d^{(r)}}
\end{equation}
subject to $\nonadap{K^{(r)} \cup J \cup V}{r}{d^{(r)}} \neq 0$. \\

Notice the combinatorial explosion that happens in this optimization problem. For a given $m$, the search space for $S^*_m$ spans over $\mathcal{O}\paran{{n-|K^{(r)}| \choose{m}}{n-m-|K^{(r)}| \choose{\sqrt{n}}}}$ subsets, each of which takes $\mathcal{O}\paran{|K^{(r)}|+m+\sqrt{n}}$ time to process. Hence, the total time it takes for the adversary in each round to solve this problem is $\mathcal{O}\paran{{n-|K^{(r)}| \choose{\tau - |K^{(r)}|}}{n-\tau \choose{\sqrt{n}}}\paran{\tau+\sqrt{n}}}$. This is highly computationally intensive, and hence, we provide experimental results against such an adversary only for small values of $n$. Note that the above problem is simply a version of the 0-1 knapsack problem and hence, is NP-HARD. 

\subsection{Experimental Results}
\begin{figure}
\centering
\includegraphics[width=0.70\textwidth]{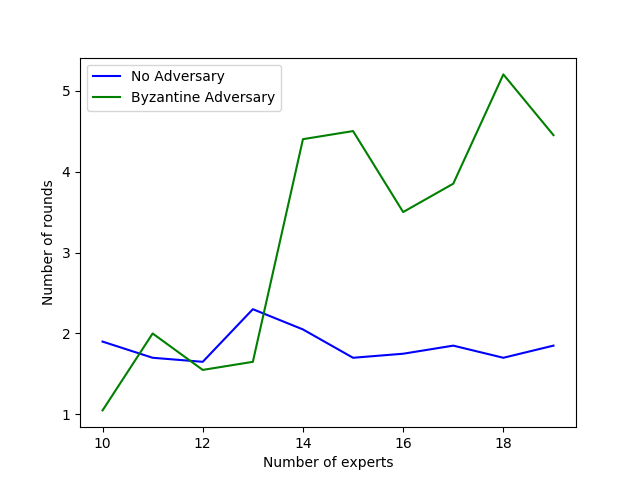}
\caption{Plot of the number of rounds before the MWU outputs in the hidden direction vs. the total number of experts. For each value of the number of experts, the corresponding number of rounds plotted is the average of running the experiment 20 times. Furthermore, the plot was obtained by keeping the fraction of the corrupted experts to be a tenth of the total number of experts.}	
\label{img:adapadv}
\end{figure}

In this section, we provide experimental results for how the MWU algorithm performs in the presence of our byzantine adversary. Our results are depicted in Fig.~\ref{img:adapadv}. Note the sharp increase in the runtime of the algorithm after $n=13$. This suggests that the adversarial strategy is indeed making the algorithm run for longer. Although the evidence is not strong, we conjecture that the number of rounds is still $\mathcal{O}(n)$, purely based on the line of best fit. 

\section{Generalization to arbitrary MWU}
\label{sec:generalization}
Having seen the two adversarial strategies for the global coin MWU, an interesting question now arises. Can we generalize the approach to an arbitrary MWU algorithm? More specifically, can we design adversarial strategies for a general purpose MWU algorithm that will help study how robust the algorithm is with respect to adversarial noise in the expert advice. We provide a brief formulation of the same here. \\

Given an MWU algorithm with $n$ experts that advice within some set $S$ in each round, assume the goal of the algorithm is to run until some formula $F$ is true. This $F$ is a mathematical statement involving the advice and the weights of the experts through the rounds. As long as $F$ is false, the algorithm runs for more rounds, and as soon as $F$ becomes true, the MWU algorithm terminates. The weight update factor $\eta$ and the payoffs for each expert are determined purely by the nature of the underlying problem and the sets $S$ and $F$. Thus, when the adversary enters the scene and has captured $K^{(r)}$ experts in round $r$, he solves a similar optimization problem as that of the adaptive adversary, but on a different function than \texttt{NON-ADAPTIVE}. The version of \texttt{NON-ADAPTIVE} that the adversary now uses is based on the formula $F$ and returns the total weight of the corrupted experts in round $r+1$ when $\lnot F$ holds, and $0$ otherwise. \\

We refer to this problem of solving the optimization instance for the adversary as the \emph{dual} of the underlying MWU algorithm. The rationale behind this name is that the adversary is himself running a computationally heavy task to compete with the MWU algorithm in order to delay the time it takes in weighing the experts so that no matter what the adversary does, eventually $F$ becomes true. In a way, the MWU algorithm acts as the adversary for this optimization problem and the two are competing with each other. An interesting question that arises here is whether all MWU algorithms are robust against their duals? Of course, we can strengthen the adversarial strategies to whatever we want, but it will be nice to explore if no matter what the adversary does, can the underlying MWU always converge to a weight distribution over the experts that is able to counter the effect of the corrupted experts. We believe the consequences of answering this question will be significant in the field of robust online learning algorithms.
\section{Conclusion} 
In this report, we provided two adversarial strategies that compete with the global coin flipping MWU algorithm to delay the time it takes to output in the hidden direction. We formulated the underlying optimization problems that the adversary needs to solve in order to ensure that he has a high share in the MWU output, and gave empirical evidence to support our claim that a global coin can be simulated in the presence of a greedy adversary in $\mathcal{O}(n)$ rounds. We then provided a brief idea of how to extend the notion of robustness against these adversarial strategies for a general MWU algorithm and highlighted the importance of determining this robustness for online algorithms in general.

\appendix
\printbibliography

\end{document}